\documentclass[a4paper,11pt]{article}
\usepackage{pos}
\usepackage{bm}

\title{Beyond Fermi-II: Intermittent Particle Acceleration by Relativistic Turbulence in Astrophysical Plasmas}
\ShortTitle{Beyond Fermi-II: Intermittent Turbulent Particle Acceleration}

\author*[a]{Anton Dmytriiev}
\author[a]{Frans van der Merwe}
\author[a]{Markus B\"ottcher}

\affiliation[a]{Centre for Space Research, \\ North-West University, Potchefstroom, 2520, South Africa}

\emailAdd{amdmame@gmail.com}
\emailAdd{frans.h.vandermerwe@gmail.com}
\emailAdd{markus.bottcher@nwu.ac.za}

\abstract{Stochastic particle acceleration in turbulent plasmas plays a key role in shaping high-energy emission from relativistic outflows, such as those in Active Galactic Nuclei (AGN) and microquasars. While traditional Fermi-II models provide a foundational framework, they often oversimplify the complex nature of realistic magnetohydrodynamic (MHD) turbulence, especially in high-amplitude ($\delta B/B_0 \sim 1$) and relativistic regimes. Recent plasma simulations for these conditions have revealed highly non-linear energization effects, such as sudden, large momentum jumps, that remain largely unexplored in astrophysical applications. We present a novel Monte Carlo framework \texttt{STRIPE} that models particle acceleration as a continuous-time random walk (CTRW), capturing both intermittent energy gains and radiative losses. The stochastic evolution of particle momenta is driven by jumps with random magnitudes determined by a distribution of magnetic-field-line velocity gradients, with synchrotron and inverse Compton cooling incorporated self-consistently. Using \texttt{STRIPE}, we explore particle acceleration under physical conditions characteristic of TeV-PeV $\gamma$-ray emitting microquasars recently identified by Large High Altitude Air Shower Observatory (LHAASO). We find that relativistic, high-amplitude turbulence naturally produces particle spectra with steep low-energy cutoffs, and hard extended power-law high-energy tails reaching tens of PeV. These features differ markedly from standard quasi-linear theory and are well suited to explaining the unexpectedly hard TeV-PeV spectra of LHAASO-detected microquasars. These results highlight turbulent acceleration in the relativistic regime as a promising mechanism for particle energization in microquasar systems, as well as potentially other extreme astrophysical environments.}

\FullConference{
High Energy Phenomena in Relativistic Outflows IX (HEPRO-IX)\\
4-8 August 2025\\
Rio de Janeiro, Brazil\\}

\newcommand{\notE}{E\mkern-12.0mu/}
\newcommand{\RE}{\mathcal{R}_{\notE}}

\begin{document}
\maketitle

\section{Introduction}
\label{sec:intro}

The stochastic, or second-order Fermi, acceleration mechanism was first proposed by Enrico Fermi in 1949 as a possible origin of cosmic rays \cite{fermi1949}. In this framework, charged particles of mass $m$ interact with non-relativistic magnetized clouds of mass $M \gg m$, moving with velocity $v_{\rm M}$ in the interstellar medium (ISM). 
%During each scattering event with a ``cloud'', particles may gain or lose a small fraction of energy. Since the probability of a collision is proportional to the relative velocity between the particle and the cloud, head-on collisions are more frequent than tail-on ones, leading to a net energy gain. 
The average energy gain per collision scales as $\Delta E/E \propto (v_{\rm M}/c)^2$, hence the term second-order Fermi acceleration.
%in contrast with first-order processes such as diffusive shock acceleration, where the gain is linear in velocity.

Subsequent developments introduced more sophisticated descriptions of stochastic acceleration based on resonant interactions between charged particles and magnetic irregularities associated with plasma waves (e.g., Alfvén waves, slow/fast magnetosonic modes). In the standard quasi-linear theory (QLT) (e.g.,\ \cite{schlickeiser1989}), magnetohydrodynamic (MHD) plasma turbulence is represented as a superposition of such plasma waves, with particles interacting resonantly with a turbulent wave spectrum.
%$W(k) \propto k^q$. The total energy in magnetic fluctuations satisfies $\int_{k_{\rm min}}^{k_{\rm max}} W(k) dk = \delta B^2/8\pi$, where $\delta B$ is the amplitude of magnetic field fluctuations and $q$ is the turbulence spectral index. 
Particle energization is then described by the Fokker–Planck equation, which governs diffusion in momentum space with a diffusion coefficient $D_{pp}$.
%$D_{pp} \propto p^q$, where $p$ is the particle momentum. 

The QLT framework remains widely used and provides accurate results in weak turbulence regimes, $\delta B/B_0 \ll 1$, where $B_0$ is the ordered magnetic field component. However, in regimes of high-amplitude, $\delta B/B_0 \sim 1$, or relativistic turbulence with Alfvén velocity $v_{\rm A} \sim c$, complex non-linear effects emerge, and the QLT theory breaks down. Analytically describing particle energization under these conditions is a significant challenge, exacerbated by the absence of a general physical theory of turbulence -- one of the major unsolved problems in modern physics.

To overcome these limitations, modern studies increasingly rely on numerical simulations, including MHD and Particle-in-Cell (PIC) approaches. 
%These simulations self-consistently follow plasma waves, velocity structures and/or individual particle dynamics in 2D/3D. MHD simulations capture large-scale dynamics over wide spatial ranges, while PIC simulations provide a fully kinetic picture of collisionless turbulence, resolving plasma-scale physics.
Recent MHD and PIC simulations of turbulence in the high-amplitude and/or relativistic regimes have revealed that particle energization is mostly dominated by non-resonant interactions with non-trivial velocity structures, rather than resonant wave-particle interactions (e.g.,\ \cite{pugliese2023}). In particular, acceleration primarily occurs at sharp bends of magnetic field lines and compression modes perpendicular to the field (e.g.,\ \cite{bresci2022}). Velocity and magnetic density fluctuations can also become highly non-Gaussian and spatially intermittent (e.g.,\ \cite{beattie2020}), leading to localized and abrupt acceleration events (e.g.,\ \cite{pezzi2022}). These effects cause particle energization to become highly intermittent, especially at small spatial scales \cite{nb2021}. Moreover, simulations indicate that acceleration can be remarkably efficient in such regimes (e.g.,\ \cite{trotta2020}).

The astrophysical relevance of these turbulence regimes is very broad. Strong turbulence ($\delta B/B_0 \sim 1$) arises when kinetic or magnetic energy is injected and rapidly dissipated, producing large-amplitude magnetic fluctuations. Relativistic turbulence, on the other hand, emerges when the plasma magnetization parameter $\sigma \gtrsim 1$, so that the Alfvén velocity approaches the speed of light, $\beta_{\rm A} = v_{\rm A}/c = \sqrt{\sigma/(1+\sigma)} \sim 1$. In relativistic outflows, such as the jets of Active Galactic Nuclei (AGN) and microquasars, these conditions are naturally fulfilled, as internal shocks, magnetic reconnection, or velocity shear between jet layers can dissipate energy efficiently.
%driving strong relativistic turbulence.

Modeling turbulent particle acceleration in these environments remains a challenging task. While MHD and PIC simulations offer a powerful means to study these processes {\it ab initio}, their high computational cost makes them impractical for exploring wide parameter ranges relevant to these systems. A promising alternative is to develop theoretical frameworks that condense and generalize the key physical insights and patterns from simulations into simplified, self-consistent and computationally efficient models. 
%Such frameworks possess predictive power, allowing to efficiently simulate particle spectra across diverse astrophysical conditions.

In this work, we focus on the framework proposed by \cite{lemoine2022}, which represents an important step in this direction, and aims to provide a physically grounded description of stochastic acceleration in collisionless, large-amplitude, and relativistic turbulence. We implement this formalism in a dedicated Monte-Carlo (MC) code \texttt{STRIPE}, extending it to include synchrotron and inverse Compton (IC) losses self-consistently. As a first astrophysical application, we explore particle acceleration in $\gamma$-ray-loud microquasar systems. Six such objects have recently been detected by LHAASO as Ultra-High-Energy (UHE; $E_{\gamma} \geq 100$ TeV) $\gamma$-ray emitters \cite{lhaaso, lhaasocygx3} (e.g.,\ SS\,433, V4641\,Sgr), exhibiting unexpectedly hard $\gamma$-ray spectra that challenge conventional acceleration scenarios. Notably, for all UHE-detected microquasars except one (Cyg X-3), the UHE emission originates not from the core region, but from extended (10--100 pc) $\gamma$-ray ``bubbles'' surrounding the systems. These large-scale structures are most likely generated by the interaction of the relativistic jet with the surrounding ISM, with high-energy particles escaping the jet termination zone and filling a much broader volume \cite{neronov2025}. The jet–ISM interaction is expected to drive strong MHD turbulence across this region, enabling efficient trapping and stochastic re-acceleration of particles. Given the inferred low particle densities, moderate magnetization, and rapid energy dissipation, the turbulence in these bubbles is likely to reach the high-amplitude ($\delta B/B_0 \sim 1$) and relativistic ($\beta_{\rm A} \sim c$) regime \cite{dmytriiev2025}. These conditions make microquasar bubbles a natural laboratory for testing strong relativistic turbulent acceleration. In this work, we investigate whether such turbulence can accelerate particles up to PeV energies in these systems, exploring also the dependence of various particle spectra signatures on key physical parameters.

\section{The Stochastic Acceleration Model}
\label{sec:lemoinemodel}

The framework developed by \cite{lemoine2022} provides a first-principles physical description of stochastic particle acceleration in strongly turbulent magnetized plasmas, explicitly accounting for the intermittent and non-Gaussian nature of turbulence. In this picture, particles are energized through random interactions with localized velocity gradients associated with turbulent magnetic field line motions. Unlike standard diffusion-based approaches, where acceleration is described by smooth average rates, this framework naturally allows for rare but intense energization events, which can dominate the high-energy tail of the particle distribution.

The evolution of particle momentum $p$ is formulated as a continuous-time random walk (CTRW), in which particles undergo discrete stochastic interactions in both momentum and time. The particle momentum is tracked in the scattering center frame $\RE$, moving with velocity $\bm{v}_E = c [\bm{E} \times \bm{B}]/B^2$, in which the electric field vanishes. In this frame, the particle energy change is directly controlled by the spatio-temporal gradients of the velocity of plasma magnetic field lines (in the laboratory frame). The instantaneous momentum evolution can be written as $\dot{p} = p \ (\Gamma_{\rm acc} + \Gamma_{||} + \Gamma_{\perp}) \equiv p \, \Gamma_{l_g}$, where the three terms represent, respectively, the effective inertial force due to acceleration of magnetic field lines, shear along the field, and transverse compression. For practical purposes, these contributions are combined into a single effective gradient $\Gamma_{l_g}$, coarse-grained on a spatial scale $l_g \sim 2\pi r_g$, where $r_g = pc/eB$ is the particle gyroradius. These gradients are highly intermittent on small scales and exhibit strongly non-Gaussian statistics.

Within the CTRW formalism, both the interaction time $\Delta t$ and the momentum increment $\Delta p$ are treated as random variables. The interaction time follows an exponential distribution, with the probability distribution function (PDF) $f_{\Delta t}(\Delta t) = (1/\tau) \ \text{exp}(-\Delta t / \tau)$, with mean $\tau = l_g(p^{\prime})/c$. The momentum change is distributed\footnote{Here, the symbol $\sim$ is referring to ``distributed as''.} as $\Delta \text{ln} \ p \sim \Gamma_{l_g} \Delta t$. When particles reach momenta $p \geq p_c$, where $p_c$ corresponds to $l_g(p_c) = l_c$, and $l_c$ is the turbulence coherence length, the mean interaction time saturates at $\tau = l_c/c$. In this regime, particles partially decouple from the turbulence, and the effective energization rate is reduced by a factor $l_c/l_g$, yielding $\dot{p} = p \ \Gamma_{l_g} \times (l_g/l_c)$.

The statistics of velocity gradients $\Gamma_{l_g}$ can be described through multifractal analysis that characterizes turbulence intermittency \cite{lemoine2022}. For practical applications, the complex multifractal PDF can be approximated by a broken power-law:
\begin{equation}
\label{eq:brokenpl}
    \mathrm{p}_{\Gamma_{l_g}} \approx \left[ 1 + \left(\frac{\Gamma_{l_g}}{\sigma(l_g)}\right)^{k_0(l_g)/k_1} \right]^{-k_1}
\end{equation}
\noindent
where $\sigma(l_g)$ is the characteristic width of the distribution, $-k_0(l_g)$ is the asymptotic power-law index at large $\Gamma_{l_g}$, and $k_1 = 3$ ensuring smoothness. 
%The parameters $\sigma$ and $k_0$ depend on the spatial scale $l_g$ as $\sigma(l_g) = \sigma(l_c) \ (l_g/l_c)^{-0.3}$, and $k_0(l_g) = 2.8 + 2(l_g/l_c)^2$ \cite{lemoine2022}, where $\sigma(l_c)$ is the width at the coherence length $l_c$. 
This PDF is additionally multiplied by an exponential cutoff $\text{exp}(-[\Gamma_{l_g} \Delta t]^2)$, 
%accounting for particles escaping the (local; small-scale) velocity gradient accelerating them, and thus 
restricting unphysically large energy gains.

A key feature of this framework is that, at small scales $l_g \ll l_c$, the PDF of $\Gamma_{l_g}$ exhibits extended power-law tails, allowing for large energy jumps over short time intervals, which may induce distinctive features in the particle spectrum at the highest energies. In contrast, at large scales $l_g \sim l_c$, the PDF becomes narrower and the acceleration process approaches the standard Fokker-Planck regime.

Energy gains and losses are determined by the sign of $\Gamma_{l_g}$. While positive and negative gradients follow the same functional form, their widths differ, leading to a net energy gain. Following \cite{lemoine2022}, the widths at the coherence length are $\sigma^{+/-}(l_c) = \sigma^{+/-}_0 \ (c/l_c) \ (\beta_a/0.41)^2$, with $\sigma^{+}_0 = 0.17$ and $\sigma^{-}_0 = 0.13$. Here, $\beta_a$ is the effective turbulent Alfvén velocity $\beta_{\rm a} = \beta_{\rm A} \ (\delta B/ B)$, with $B = \sqrt{B_0^2 + \delta B^2}$. The ``background'' Alfvén velocity is $\beta_{\rm A} = \sqrt{\sigma/(1+\sigma)}$. 

All statistical distributions and scaling relations in this model were calibrated by \cite{lemoine2022} using numerical MHD simulations of driven incompressible turbulence, ensuring scalability. The main physical control parameters of the model are therefore the magnetic field strength $B$, the turbulence coherence length $l_c$, and the effective turbulent Alfvén velocity $\beta_a$.

\section{The \texttt{STRIPE} code}

We implement the framework by \cite{lemoine2022} in a dedicated Monte Carlo (MC) code \texttt{STRIPE} (Strong-Turbulence Relativistic Intermittent Particle Energization). For astrophysical applications, we extend this formalism to include radiative losses of particles due to synchrotron and/or IC emission, as well as particle escape from the large-scale acceleration region. The escape process is modeled as energy-independent. This choice reflects the current lack of a robust theoretical prescription for energy-dependent spatial particle transport in strongly intermittent turbulence, which leads us to adopt a minimal parameterization with a single characteristic escape time-scale $t_{\rm esc}$. While energy-dependent escape may appreciably affect the shape of particle spectra, its inclusion would require additional poorly constrained assumptions and is deferred to future research. In its base configuration, the code assumes single-pulse (instantaneous) and mono-energetic injection of electrons with an initial Lorentz factor $\gamma_0$.

The main input parameters of \texttt{STRIPE} are: magnetic field $B$, turbulence coherence length $l_c$, effective turbulent Alfvén velocity $\beta_a$, injection Lorentz factor $\gamma_0$, Compton dominance $CD$, simulation start and end times $t_{\rm s}$ and $t_{\rm f}$ (in units $l_c/c$), number of logarithmic time grid points $n_{\rm tp}$, initial number of particles $N_{\rm p}$ and the escape time-scale $t_{\rm esc}$ (in units $l_c/c$).

To incorporate the radiative losses, we derive an expression describing momentum change over a single stochastic interaction. Although particle energization arises from discrete stochastic interactions, it is treated here in an effective quasi-continuous form through the relation $\dot{p} = p \Gamma_{l_g}$, following \cite{lemoine2022}. Radiative losses, on the other hand, are intrinsically continuous processes and are treated as such. The radiative cooling rate in the ultra-relativistic regime (particle Lorentz factor $\gamma \gg 1$) is given by $\dot{\gamma} = - b_{\rm c} \gamma^2$, where $b_{\rm c} = \frac{4 \sigma_{\rm T}}{3 m_{\rm e} c} \ U_B \ (1 + CD)$. Here $U_B = \frac{B^2}{8\pi}$, $CD = U_{\rm rad}/U_B$ is the Compton dominance, and $U_{\rm rad}$ is the energy density of the incident photon field. This form is valid in the Thomson regime of IC scattering. In the Klein–Nishina (KN) regime, we apply a correction factor $q_{\rm KN}$ to $U_{\rm rad}$, yielding an effective photon energy density $U_{\rm rad,eff} = q_{\rm KN}\,U_{\rm rad}$, with $q_{\rm KN}$ being computed using the analytical approximation of \cite{moderski2005}. Rewriting the radiative loss term in terms of the particle momentum $p = \gamma m_{\rm e} c$ and combining it with the stochastic acceleration term, we obtain the full momentum evolution equation:
\begin{equation}
    \frac{dp}{dt} \ = \ p \Gamma_{l_g} \ - \ \frac{b_{\rm c}}{m_{\rm e} c} p^2
\end{equation}
\noindent
Solving this differential equation yields the momentum change over a single interaction time $\Delta t$:
\begin{equation}
\label{eq:momen_evol_cool}
    p = \frac{p^{\prime} \text{exp}(\Gamma_{l_g} \Delta t)}{1 \ + \ \varepsilon \gamma^{\prime} \left[\text{exp}(\Gamma_{l_g} \Delta t) - 1\right]}
\end{equation}
\noindent
where $\gamma^{\prime} = p^{\prime}/ m_{\rm e} c$ is the initial Lorentz factor of the particle, and $\varepsilon = b_{\rm c}/\Gamma_{l_g}$. Note that $\Gamma_{l_g}$ can take either positive or negative sign depending on whether the interaction leads to energy gain or loss. In the absence of radiative cooling ($b_{\rm c}=0$), Eq.~\ref{eq:momen_evol_cool} reduces to the simple form:
\begin{equation}
\label{eq:momen_evol_nocool}
    p = p^{\prime} \ \text{exp}[\Gamma_{l_g} \Delta t]
\end{equation}
\noindent
The code is written in \texttt{C} and parallelized with the Message Passing Interface (MPI), distributing $N_{\rm p}$ particles across available CPU cores. After initialization, $N_{\rm p}$ electrons with Lorentz factor $\gamma_0$ are evolved independently in time, which corresponds to the mono-energetic single pulse injection (SPI) mode. During each interaction step, the code randomly draws ({\it i}) the interaction time $\Delta t$ from the exponential distribution defined in \ref{sec:lemoinemodel}, ({\it ii}) the sign of the interaction (energy gain or loss) according to their relative probabilities, and ({\it iii}) the velocity gradient $\Gamma_{l_g}$ from Eq.~\ref{eq:brokenpl}. Using these random variables, the particle momentum is updated according to Eq.~\ref{eq:momen_evol_cool} (or Eq.~\ref{eq:momen_evol_nocool} without cooling), and the elapsed time is advanced by $\Delta t$. For each particle, a single ``survival'' time is drawn at initialization from an exponential distribution around $t_{\rm esc}$. The particle is removed from the simulation when its elapsed time exceeds the survival time. Particle evolution also terminates if the total elapsed time reaches the final simulation time $t_{\rm f}$. The momentum evolution of each particle is recorded at all time grid points, allowing the construction of time-dependent electron spectra.

While the default configuration assumes SPI mode, other injection time profiles can be modeled in post-processing through a dedicated module. The most common and astrophysically relevant case is continuous injection (CI) at a constant rate $Q$. In this case, to obtain the corresponding electron spectra $N_{e,{\rm CI}}(\gamma,t)$, we numerically convolve the SPI spectrum $N_{e,{\rm SPI}}(\gamma,t)$ (serving as the Green's function solution) with the Heaviside function $\Theta(t-t_{\rm s})$. Under our assumption of mono-energetic injection, we verified that the resulting asymptotic particle spectra are largely insensitive to the exact choice of $\gamma_0$. This is because, for sufficiently long evolution times, particles rapidly lose memory of their initial energy due to the stochastic nature of the acceleration process. The only noticeable effect is that, in the CI case, mono-energetic injection introduces a low-energy cutoff around $\gamma_0$. More complex injection prescriptions (e.g.,\ power-law or Maxwellian distributions), which may be relevant in re-acceleration scenarios, are expected to mainly affect the low-energy part of the spectrum and are left for future work.

%\begin{equation}
%   N_{e,\mathrm{CI}}(\gamma, t_*) = \sum_{i=1}^{M_{t_*}} \frac{1}{N_{\rm p}} N_{e,\mathrm{SPI}}(\gamma,t_i) \times Q \Delta t_i
%\end{equation}

%where $\Delta t_i$ and $t_i$ denote the width and time value of the $i$th grid step, and $M_{t_*}$ corresponds to the index for which $t_{i=M_{t_*}} = t_*$. The total number of particles integrated over $N_{e,{\rm CI}}(\gamma, t_*)$ asymptotically approaches $N_{p,{\rm CI}} = Q t_{\rm esc}$ as $t_* \to \infty$, as expected.

The simulated spectra are computed in the scattering-center frame and must be transformed to the observer's (laboratory) frame. This is done by numerical convolution with a boosting kernel parameterized by the effective turbulent velocity $\beta_a$, following the procedure detailed in \cite{lemoine2022}.

\section{Application to Microquasars}

As outlined in Section ~\ref{sec:intro}, TeV–PeV–emitting microquasars host large multi-parsec-scale ``bubbles'' in which particles can be efficiently re-accelerated by large-amplitude relativistic turbulence. The re-accelerated electrons produce X-ray synchrotron radiation, as well as IC upscatter Cosmic Microwave Background (CMB) photons to generate TeV–PeV $\gamma$-ray emission. Various modeling efforts constrain the bubble magnetic field to be of the order $B \sim 10 \mu$G \cite{wan2025, dmytriiev2025}, while the turbulence coherence length is expected to be $l_c \sim 1$ pc, given the spatial dimensions of the bubbles. We therefore adopt these as benchmark parameters and explore the surrounding parameter space in the context of turbulent particle acceleration. The Compton dominance for CMB seed photons in the frame of the bubble is $CD = U_{\rm CMB} \Gamma_{\rm b}^2 / U_B$, where $\Gamma_{\rm b}=(1-\beta_{\rm b}^2)^{-1/2}$ is the bulk Lorentz factor and $U_{\rm CMB} \approx 4.16 \times 10^{-13} \ {\rm erg \ cm^{-3}}$. For the weakly to mildly relativistic outflows expected here ($\beta_{\rm b}\lesssim 0.1$) and for $B\gtrsim 10 \mu$G, we find $CD \lesssim 0.1$. IC losses are therefore subdominant, and for simplicity we neglect them completely, adopting pure synchrotron cooling.

\begin{figure}
    \centering
    \includegraphics[height=0.34\linewidth]{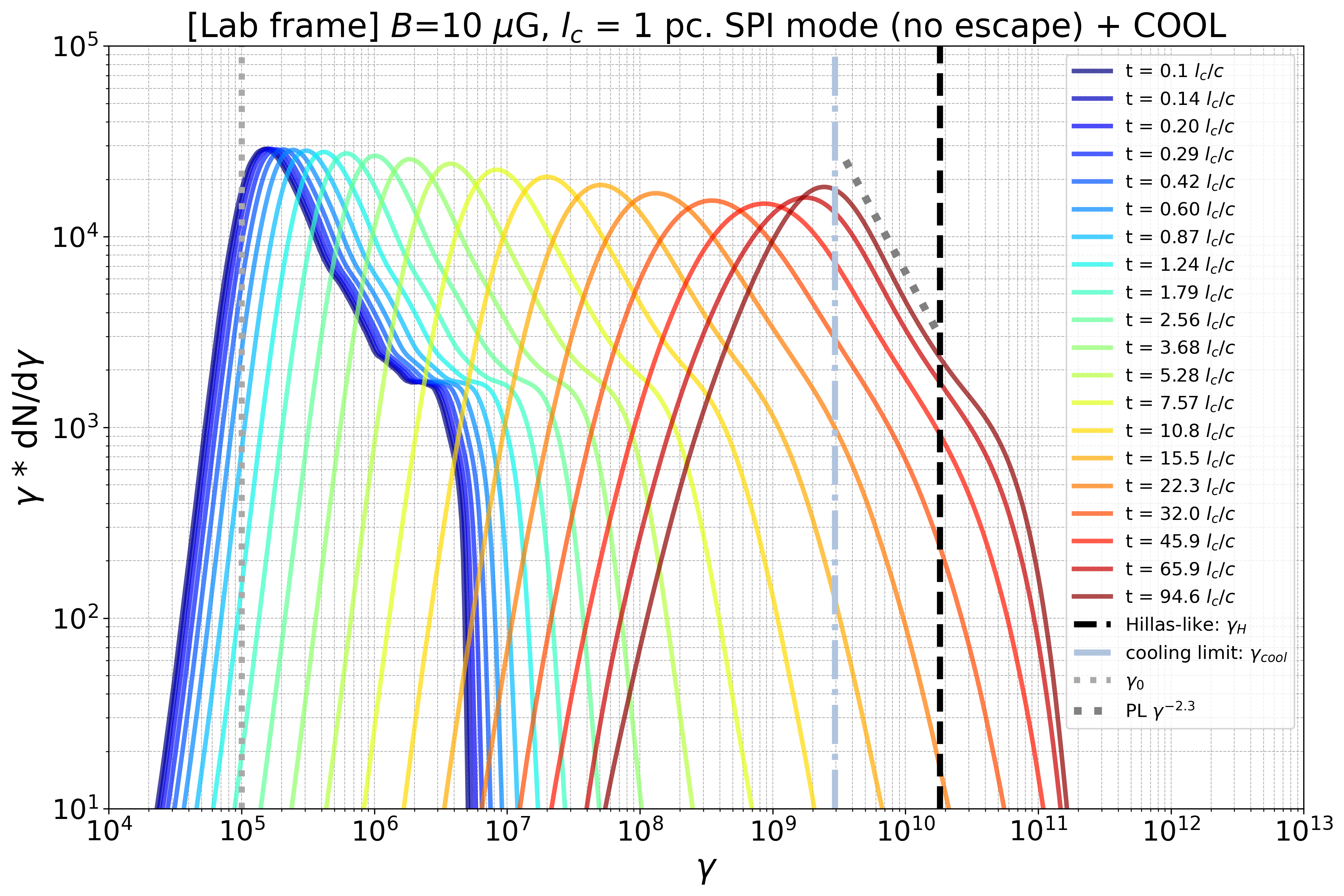} \includegraphics[height=0.34\linewidth]{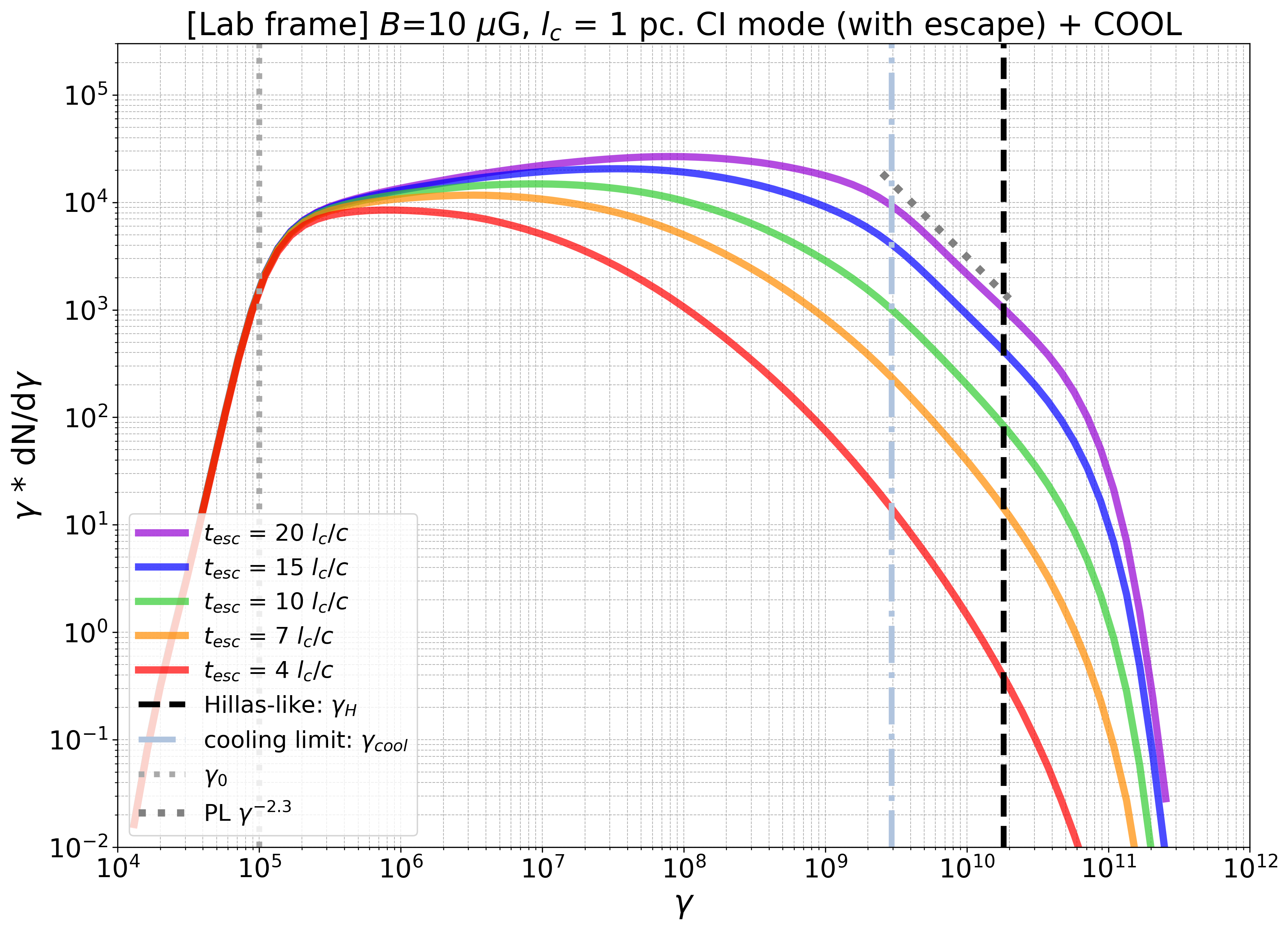}
    \caption{Electron spectra for the baseline parameter set $B = 10 \mu$G, $l_c = 1$ pc, $\beta_a = 0.41$, $\gamma_0 = 10^5$. Left: time evolution of the electron spectrum in the SPI mode (with synchrotron cooling included). Right: corresponding steady-state electron spectra in the CI mode for different values of $t_{\rm esc}$.}
    \label{fig:el_spec_baseline}
\end{figure}

We consider two modes of particle injection: the SPI and CI modes. In all SPI-mode runs particle escape is disabled, whereas in the CI mode  we include escape with different $t_{\rm esc}$ values. Using \texttt{STRIPE}, we compute electron spectra for the baseline set with $B = 10 \mu$G, $l_c = 1$ pc, $\beta_a = 0.41$, $\gamma_0 = 10^5$ for both SPI and CI modes (Fig.~\ref{fig:el_spec_baseline}).

\begin{figure}
    \centering
    \includegraphics[height=0.33\linewidth]{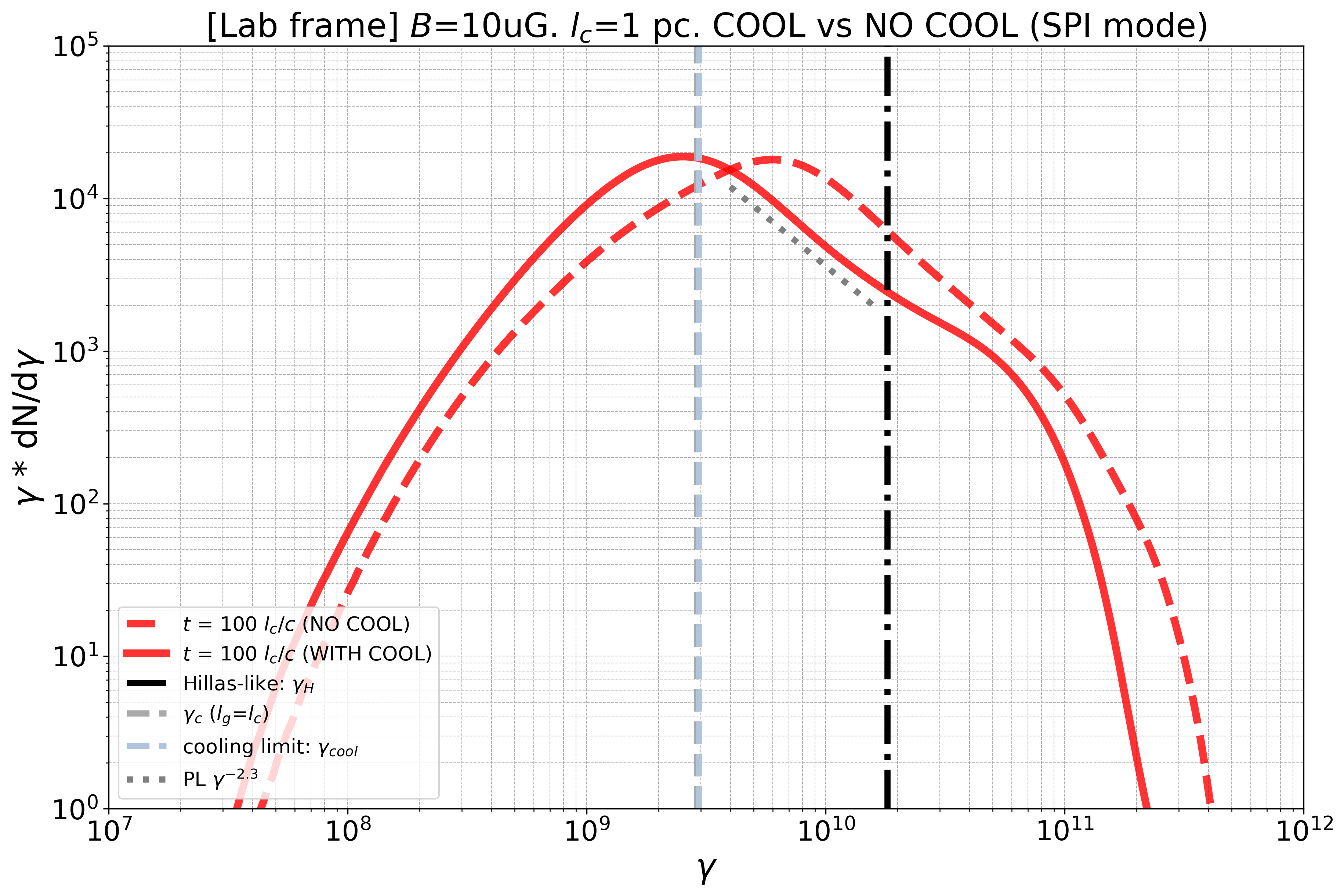} \includegraphics[height=0.33\linewidth]{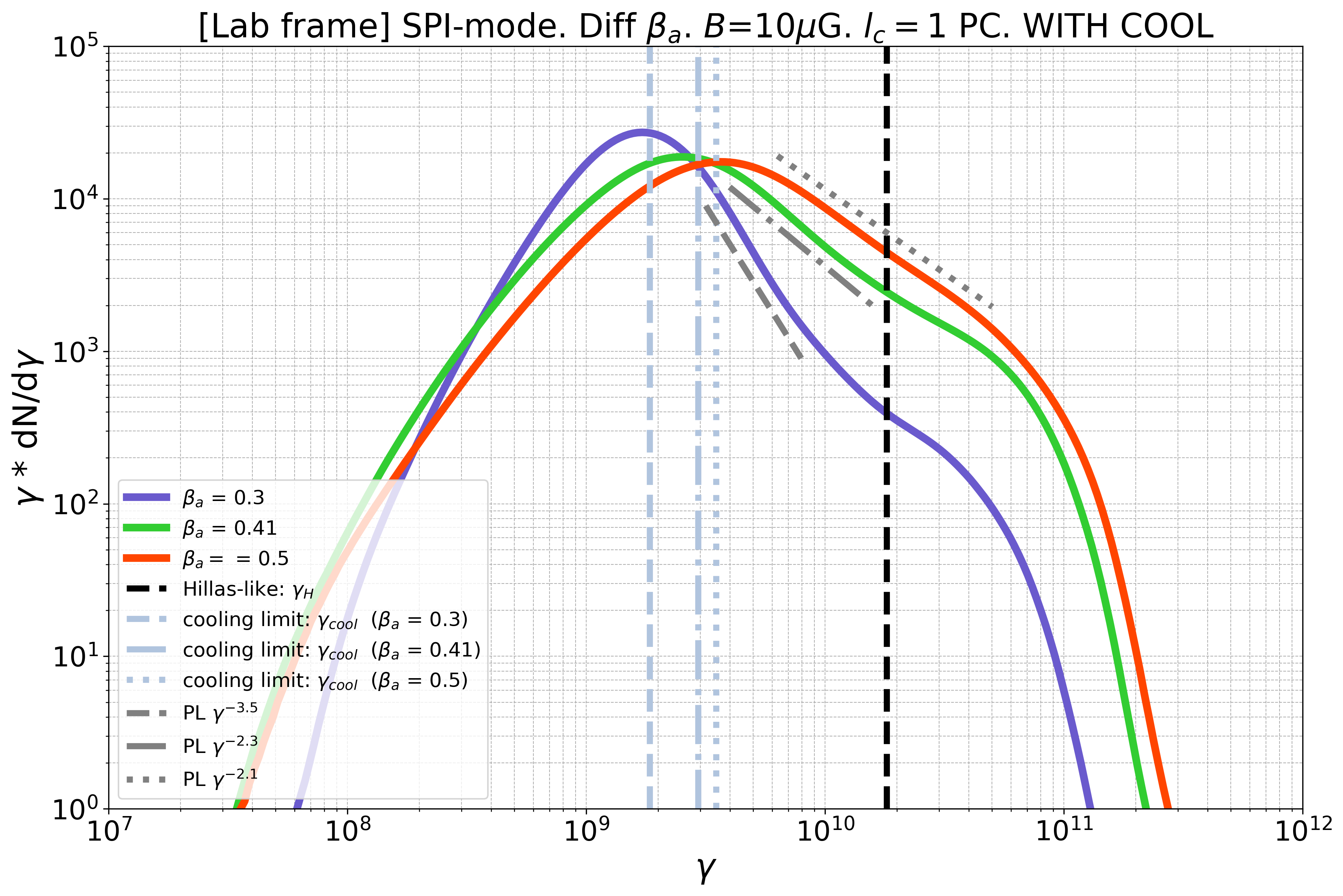}
    \includegraphics[height=0.33\linewidth]{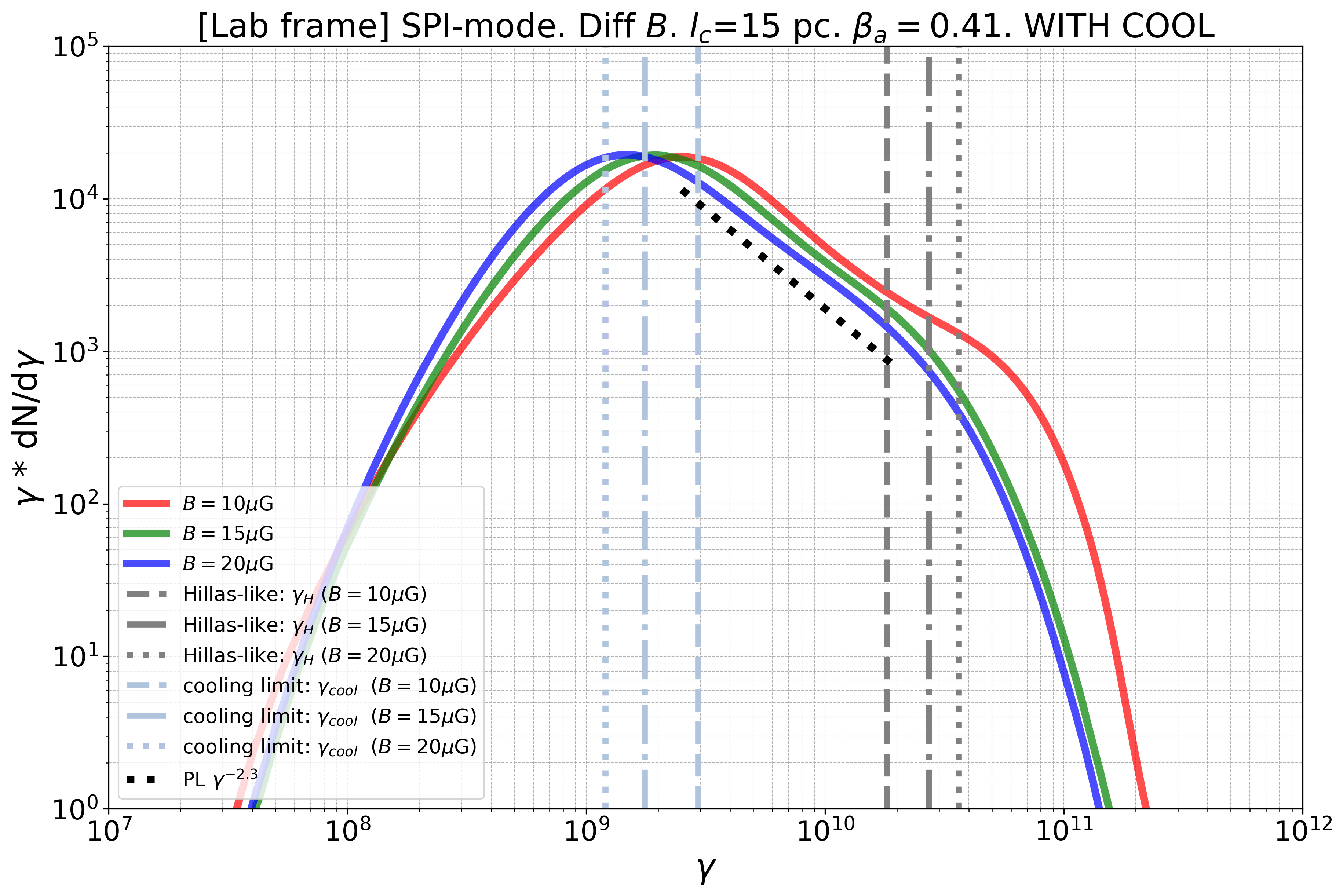}
    \includegraphics[height=0.33\linewidth]{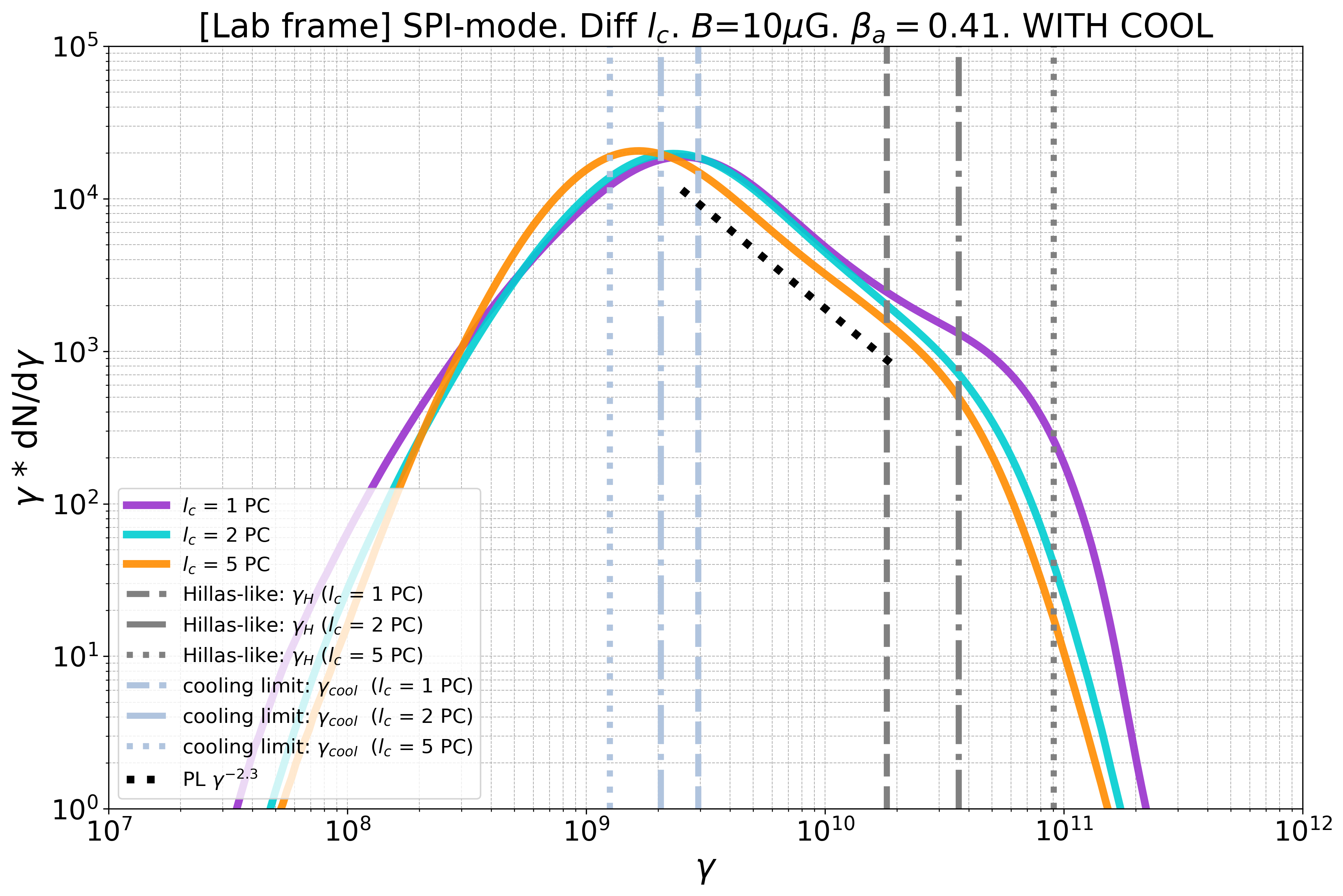}
    \caption{Asymptotic electron spectra in the SPI mode for the baseline parameter set $B = 10 \mu$G, $l_c = 1$ pc, $\beta_a = 0.41$, $\gamma_0 = 10^5$. Top left: comparison of spectra with synchrotron cooling enabled and disabled. Top right: dependence on turbulent Alfvén speed $\beta_a$. Bottom left: dependence on magnetic field strength $B$ (cooling enabled). Bottom right: dependence on coherence length $l_c$.}
    \label{fig:effect_cool_B_lc_betaa}
\end{figure}

We then investigate the dependence of the asymptotic SPI-mode spectrum on the main physical parameters. Fig.~\ref{fig:effect_cool_B_lc_betaa} shows the resulting spectra and their dependence on radiative losses and key turbulence parameters, namely the magnetic field strength $B$, coherence length $l_c$, and turbulent Alfvén speed $\beta_a$. Across all parameter variations, the spectra share the same characteristic shape: a steep low-energy cutoff, a relatively narrow spectral peak, and an extended high-energy power-law (PL) tail that terminates in a sharp cutoff at the highest energies. Without cooling, the spectral peak forms near the Lorentz factor $\gamma_c$ at which $l_g(\gamma_c)=l_c$, beyond which particles start to decouple from the turbulence. When cooling is included, the spectral peak instead settles at the Lorentz factor $\gamma_{\rm cool}$ at which synchrotron losses balance the effective acceleration rate, $b_c \gamma_{\rm cool} \simeq \langle\Gamma_{l_g(\gamma_{\rm cool})}\rangle$. Importantly, the high-energy PL tail extends one to two orders of magnitude beyond $\gamma_{\rm cool}$ -- a characteristic feature of the turbulence intermittency. The high-energy cutoff occurs close to the Hillas-like limit, $\gamma_{\rm H} \ m_{\rm e} c^2 \simeq eBl_c$.

The slope of the high-energy PL tail, $\alpha_{\rm HE}$, is found to be largely insensitive to $B$ and $l_c$, but strongly dependent on $\beta_a$, with larger $\beta_a$ yielding systematically harder spectra: $\alpha_{\rm HE} \simeq$ 3.5, 2.3 and 2.1 for $\beta_a$ = 0.3, 0.41 and 0.5. Notably, the PL-tail index remains unchanged when radiative losses are included, indicating that the tail is shaped predominantly by intermittent turbulent energization.

\begin{figure}
    \centering
    \includegraphics[width=0.6\linewidth]{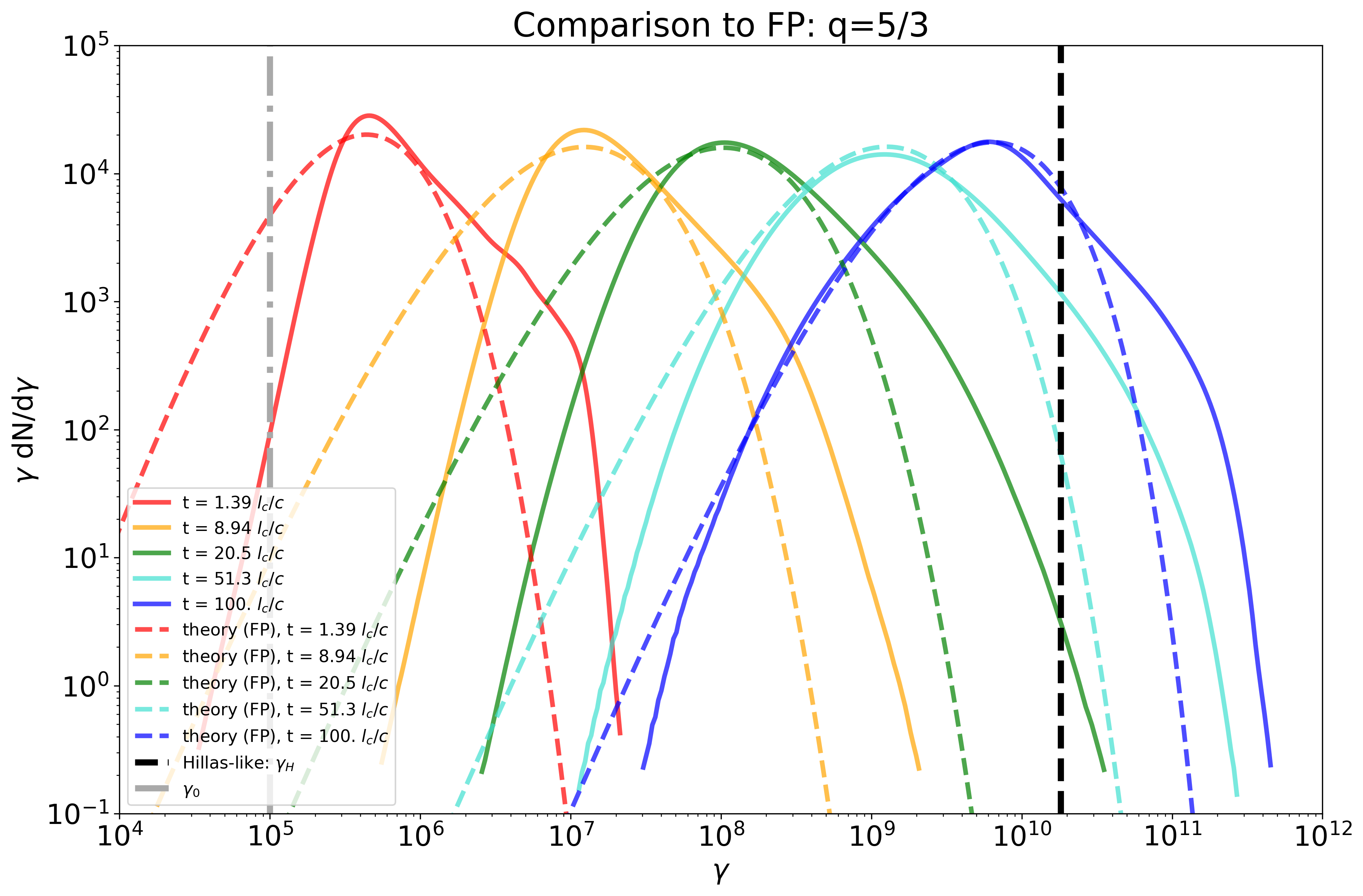}
    \caption{Comparison of the MC-simulated electron spectra for baseline parameter set with cooling disabled (solid lines) to corresponding solutions of Fokker-Planck equation at different evolution times (dashed lines).}
    \label{fig:fokker_planck}
\end{figure}

To place these results in context, we compare the simulated SPI-mode spectra (without cooling) to analytic Fokker–Planck (FP) solutions with zero escape from \cite{becker2006}. Although FP theory formally breaks down for strong turbulence, the comparison (see Fig.~\ref{fig:fokker_planck}) highlights the qualitative differences: all MC-simulated spectra show steeper low-energy cutoffs, while also exhibit hard extended power-law tails (rather than log-parabolic curved shape), and also display narrower peaks at early evolution times. These effects arise from the highly intermittent nature of the acceleration process, where particles experience sporadic large jumps rather than diffusing gradually in momentum space.

For the microquasar parameters explored here, turbulent acceleration is able to energize electrons to maximum energies of $\sim 50$ PeV, while producing high-energy tails with indexes $\alpha_{\rm HE} \lesssim 2.3$ for $\beta_a \gtrsim 0.41$. These properties are sufficient to explain the hard TeV–PeV spectra reported by LHAASO. Finally, we applied the complete acceleration and radiation framework (combining \texttt{STRIPE} with the synchrotron–IC emission module) to the microquasar V4641\,Sgr. As demonstrated in \cite{dmytriiev2025}, this model successfully reproduces the observed broadband SED of the source.

\section{Conclusions}

We have implemented the relativistic large-amplitude turbulence acceleration framework of \cite{lemoine2022} in a MC code \texttt{STRIPE} and extended it to include synchrotron (and optionally IC) cooling. Using this tool, we explored the particle acceleration signatures expected in large-scale microquasar environments, where jet–ISM interactions likely drive strong relativistic turbulence. For benchmark parameters motivated by LHAASO-detected systems, the SPI-mode electron spectra display relatively narrow spectral peaks, hard extended high-energy power-law tails with indexes in the range 2.1--3.5, and maximum electron energies reaching tens of PeV. These characteristics align with the requirements for explaining the observed TeV–PeV emission from LHAASO-detected microquasars. Our results suggest that turbulent re-acceleration in the relativistic large-amplitude regime provides a viable and promising mechanism for shaping high-energy spectra in these systems. Further work will focus on applications to other sources, such as extreme blazars, etc.

\section{Acknowledgments}

We thank M.~Lemoine and L.~Comisso for useful discussions. The authors acknowledge support from the Department of Science, Technology and Innovation, and the National Research Foundation of South Africa through the South African Gamma-Ray Astronomy Programme (SA-GAMMA). FvdM is supported with the funding from the National Astrophysics and Space Science Programme (NASSP) of South Africa.

\end{document}